\documentclass[fleqn,12pt,twoside]{article}
\usepackage[headings]{espcrc1}

\readRCS
$Id: espcrc1.tex,v 1.2 2004/02/24 11:22:11 spepping Exp $
\usepackage{graphicx}
\usepackage[figuresright]{rotating}

\newcommand{\AmS}{{\protect\the\textfont2
  A\kern-.1667em\lower.5ex\hbox{M}\kern-.125emS}}

\title{Bulk hadron production at high rapidities}

\author{G. I. Veres\address[ELTE]{E\"otv\"os Lor\'and University,
        Department of Atomic Physics\\ 
        P\'azm\'any P\'eter s\'et\'any 1/A, H-1117, Budapest, Hungary}$^{\rm ,}$\address[MIT]{Massachusetts Institute of Technology, 
        Laboratory for Nuclear Science\\
        77 Massachusetts Avenue, Bldg. 24-422, Cambridge, MA 02139, USA}}       

\runtitle{Bulk hadron production at high rapidities}
\runauthor{G. I. Veres}
\begin{document}

\maketitle
\begin{abstract}
Recent experimental observations on the `bulk'
features of particle production at high (pseudo)rapidities will be 
reviewed. This kinematic 
region is of interest mostly because of its relevance to the
theoretical description of initial 
state effects of nuclei at ultra-relativistic energies. 
Measurements of the charged hadron multiplicity density as well as the 
pseudorapidity dependence of the elliptic and directed flow exhibit a 
remarkable scaling property as a function of collision energy. This scaling 
seems to hold for pions and even photons and J/$\Psi$-s, but is violated 
for protons. The special role of baryons will be discussed using 
selected results on nuclear transparency and baryon stopping.
\end{abstract}
\section{INTRODUCTION}

\subsection{Motivation}

The focus of research in relativistic heavy ion collisions initially has 
been toward the mid-rapidity region. This is justified 
in the context of the search for 
the new phase of QCD at high temperatures (energy density), the Quark 
Gluon Plasma, the emergence of which was expected at 
midrapidity. In addition, an extended Lorentz-invariant central plateau was 
predicted in the (charged) particle rapidity 
distribution \cite{bjorken}. Most experiments have been designed with good 
midrapidity-coverage in mind. At earlier fixed target experiments, 
particle detection closer to beam (target) rapidity was less difficult 
than at high energy colliders.

In heavy ion collisions, the existence of a
boost-invariant {\it central} region is not confirmed 
experimentally. On the other hand,
several observables in the {\it high rapidity} region are found 
to scale in a simple way as a function of energy. Furthermore, most of the 
baryons are emitted at high rapidity, and their energy 
loss is of primary interest, since it is related to the 
created energy density. The total number of produced charged particles 
per participant nucleon ($\rm{N}_{\rm part}$) pair does not depend 
on the centrality of the heavy ion collision. There is a remarkable 
balance of the centrality dependent midrapidity and high rapidity 
particle yields \cite{phoboswhite}, suggesting a deeper 
connection between these two regions.

At RHIC, a suppression of high-$p_T$ particles at high rapidities is 
observed in d+Au col\-li\-si\-ons \cite{brahmscgc}, which is 
interpreted as the saturation of the gluon density of the 
ultra-relativistic Au nucleus at low Bjorken-x values 
\cite{gyulassy}.
Measurements at high rapidity are necessary to reach small enough $x$ 
values 
but still have a large enough momentum transfer (${\rm Q^2}$) for this 
interpretation to apply. However, it is not easy to 
disentangle which physical processes (Cronin-effect, shadowing, 
energy loss, initial state gluon saturation, fragmentation of partons at 
high rapidities, etc.) influence the observed final hadron
yields.

There is a large amount of data available on
collisions of simpler systems (p+p, p+A)
partially still under analysis. 
Baryon stopping and nuclear transparency is a large field
of study in itself, and is also related to basic questions 
about the carriers of baryon number.
Parton energy loss in the QGP and in cold nuclei are at the 
center of attention ever since the discovery of the suppression of high $p_T$ 
particle yields (`jet quenching') \cite{jetq}. The observed 
baryon `anomaly' (unexpectedly high $p/\pi^+$ and $\overline{p}/\pi^-$ 
ratios
at high $p_T$) \cite{banomaly} suggests another connection 
between dynamics of baryons at midrapidity in the transverse direction, 
and baryon transport in the longitudinal direction \cite{vasile}.
A non-exhaustive set of measurements will be discussed here, to 
illustrate some of the thoughts outlined above.

\subsection{Kinematics of saturation physics}

\begin{figure}[t]
\centerline{
\vspace*{-10mm}
\includegraphics[width=95mm]{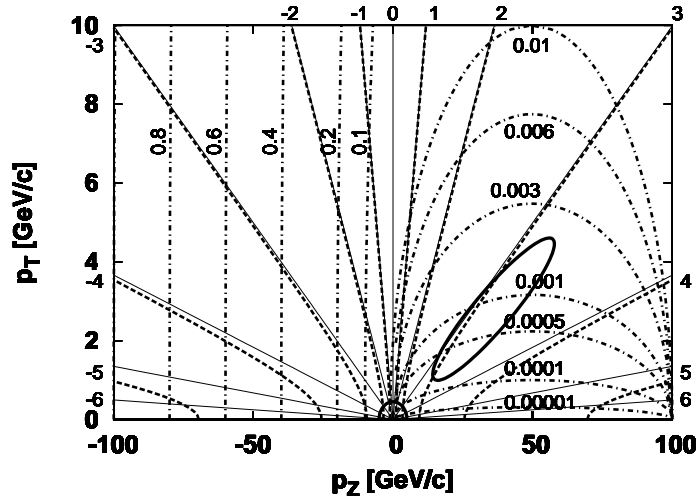}}
\caption{\label{kinematics}Constant $y$ (dashed lines, for proton mass),
$\eta$ (thin solid lines) and $x_{\rm Bj}$ (dash-dotted lines) 
contours in the $p_T$ vs. $p_Z$ plane for the 
highest RHIC energy. (See the text for details.)
The PHOBOS acceptance for very low $p_T$
\cite{adamtalk} (low-$x_{\rm Bj}$ and low $\rm Q^2$), and part of the BRAHMS acceptance at high $\eta$
(low-$x_{\rm Bj}$ and higher $\rm Q^2$) are shown by ellipses.}
\vspace*{-6.5mm}
\end{figure}

Various kinematic variables are used to study high rapidity phenomena
and saturation physics, where the largest depletion of gluon density is
expected at low $x_{\rm Bj}$ and high ${\rm Q^2}$; experiments have to
focus on that region.
Fig. \ref{kinematics} shows constant rapidity ($y$, dashed lines, for proton mass), 
and pseudorapidity ($\eta$, thin solid lines) lines of final state particles
in the transverse ($p_T$) vs. longitudinal momentum 
($p_Z$) plane for the highest RHIC energy (100 GeV/nucleon colliding beams).
As an illustration, constant Bjorken-x ($x_{\rm Bj}$) lines of partons probed in the Au nucleus are also 
plotted (dash-dotted lines, ignoring mass) for a d+Au collision (with the
Au nucleus moving from right to left)\footnote{We use a great simplification here: we think of the $d$ 
projectile as the incoming `lepton' and the final state particle as the outgoing `lepton', and borrow the 
definition of $x_{\rm Bj}$ from Deep Inelastic Scattering.}.
A fourth variable, Feynman-x (not shown) is defined as 
$x_F=2p_Z/\sqrt{s_{_{\rm NN}}}$, thus $x_F\approx x_{\rm Bj}$ for large $x_{\rm Bj}$.

\subsection{Experimental capabilities}
At RHIC, several detectors cover the high rapidity region. 
The STAR experiment has the Photon Multiplicity 
Detector \cite{starpmd} ($2.3<\eta<3.8$),
Forward TPC \cite{starftpc} ($2.5<|\eta|<4.0$), 
Forward $\pi^0$ Detector \cite{starfpd} ($3.3<\eta<4.1$) and the planned 
Forward Meson Spectrometer ($2.5<\eta<4.0$). The PHENIX experiment can use 
the muon stations to measure hadron yields at high rapidities 
($1.4<|\eta|<2.2$), using the decay length dependence of the 
$K^\pm\rightarrow\mu^\pm\nu_\mu(\overline{\nu_\mu})$
and 
$\pi^\pm\rightarrow\mu^\pm\nu_\mu(\overline{\nu_\mu})$ 
decays, or observing punch-through hardons \cite{phenixmuon}.
The BRAHMS spectrometers cover a large pseudorapidity region ($\eta<3.5$) with particle-ID 
using Ring Imaging \v{C}erenkov Detectors \cite{brahmsexp}. Finally, the largest $\eta$ acceptance 
(although with no particle identification) is offered
by the PHOBOS Multiplicity Array $|\eta|<5.4$ \cite{phobosnim}.
As examples, the PHOBOS acceptance for very low $p_T$ 
\cite{adamtalk} (low-$x_{\rm Bj}$ and low $\rm Q^2$), and the BRAHMS acceptance at high $\eta$ 
(low-$x_{\rm Bj}$ and higher $\rm Q^2$) are illustrated in Fig. \ref{kinematics}. 
The low $x_{\rm Bj}$ and high $\rm Q^2$ region is at high $\eta$ and high $p_T$.
At RHIC, this kinematic window is rather narrow, however, a large momentum space window to study 
low-$x_{\rm Bj}$ physics will be available, even at midrapidity, to the future LHC experiments.
The CMS experiment plans to cover 
an extensive range of pseudorapidity ($|\eta|<6.9$) with its calorimeters \cite{murraytalk}.

\section{ENERGY DEPENDENCE OF PARTICLE PRODUCTION AT HIGH $\eta$}

\subsection{Inclusive charged particle spectra}

\begin{figure}[t]
\centerline{
\vspace*{-10mm}
\includegraphics[width=67mm]{pplimfrag.eps}
\hspace{1mm}
\includegraphics[width=71mm]{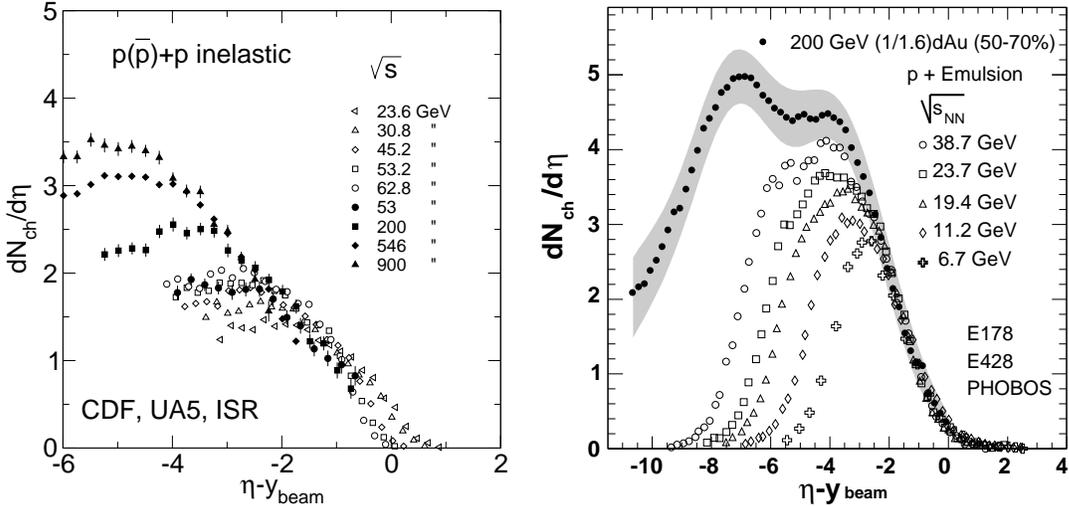}} 
\caption{\label{pplimfrag} Pseudorapidity density distributions of charged particles emitted 
in $p(\overline{p})+p$ (left panel) and d+Au, p+Emulsion (right panel) collisions at various 
energies as a function of the $\eta'=\eta-y_{\rm beam}$ variable
\cite{pplimfrag,oldpa,phobosdAumult,brahmsdau}. For details of the normalization of the data in the right 
panel see \cite{phobosdAumult}.}
\vspace*{-6.5mm}
\end{figure}

One of the simple ways to study particle production at high rapidity is to compare
charged particle $\eta$ distributions measured at various collision energies.
The $\eta'$=$\eta$-$y_{\rm beam}$ variable is used. It is measurable without 
particle identification and approximately transforms the distributions to the rest frame
of one of the colliding particles/ions.

In proton-proton collisions, the so called {\it limiting fragmentation}
\cite{benecke} is seen
from a few GeV to 900 GeV collision energy (e.g. at CDF, UA5 and ISR 
\cite{pplimfrag}). 
The left panel of Fig. \ref{pplimfrag} shows that the charged particle pseudorapidity
density distributions are approximately independent of 
collision energy over a large range of $\eta'$, which grows with $\sqrt{s}$.

As the right panel of Fig. \ref{pplimfrag} shows, similar scaling was observed 
for p+A collisions between $\sqrt{s_{_{\rm NN}}}=6.7$ and
38.7 GeV \cite{oldpa}, and more recently in d+Au collisions at $\sqrt{s_{_{\rm NN}}}=200$ 
GeV \cite{phobosdAumult,brahmsdau}. 
The observed energy-independence of 
these distributions holds both for the projectile and for the target reference frames, i.e. for
$\eta\pm y_{\rm beam}$ \cite{phobosdAumult}.

\begin{figure}[t]
\centerline{
\vspace*{-8mm}
\includegraphics[angle=-90,width=82mm]{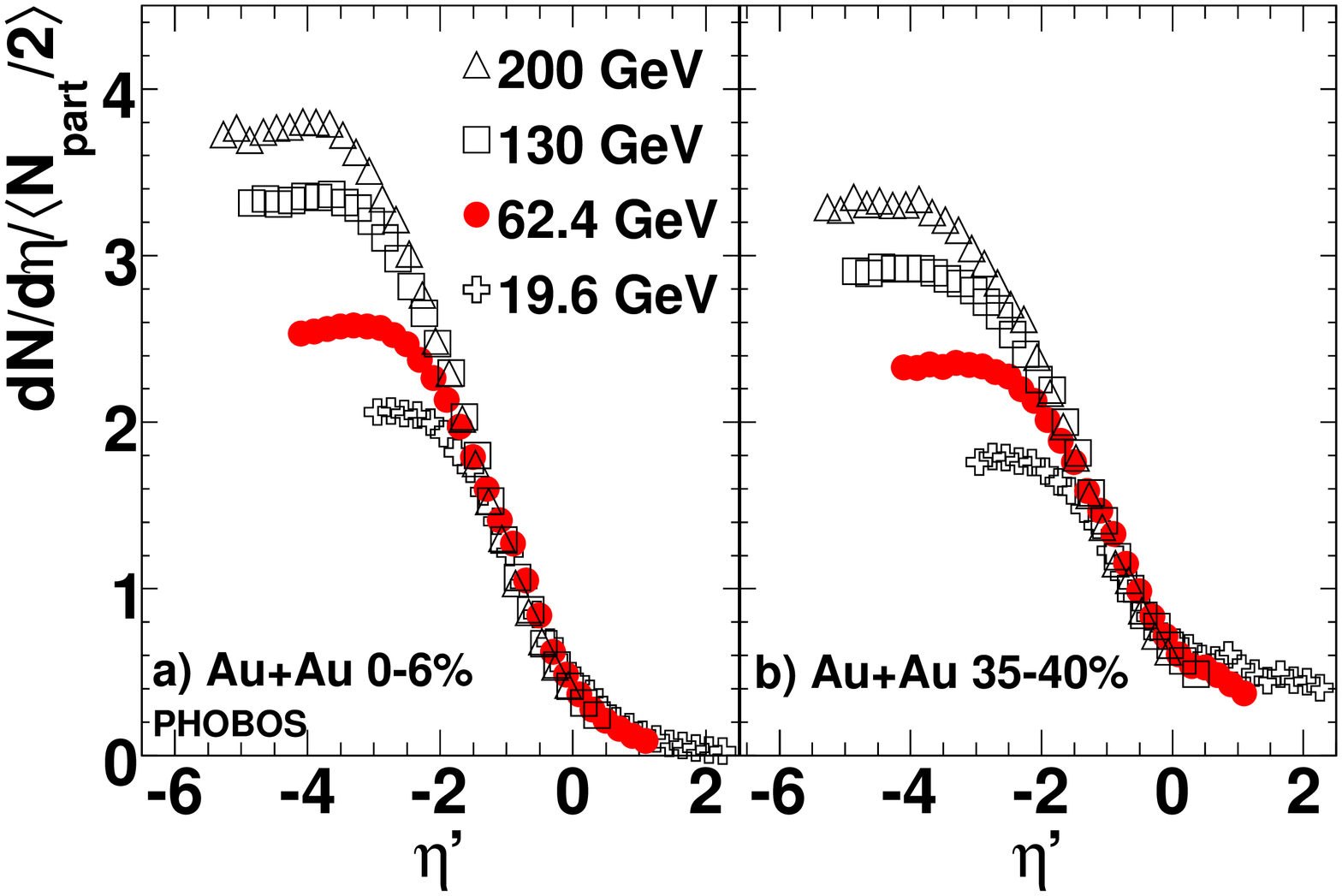}
\includegraphics[angle=-90,width=77mm]{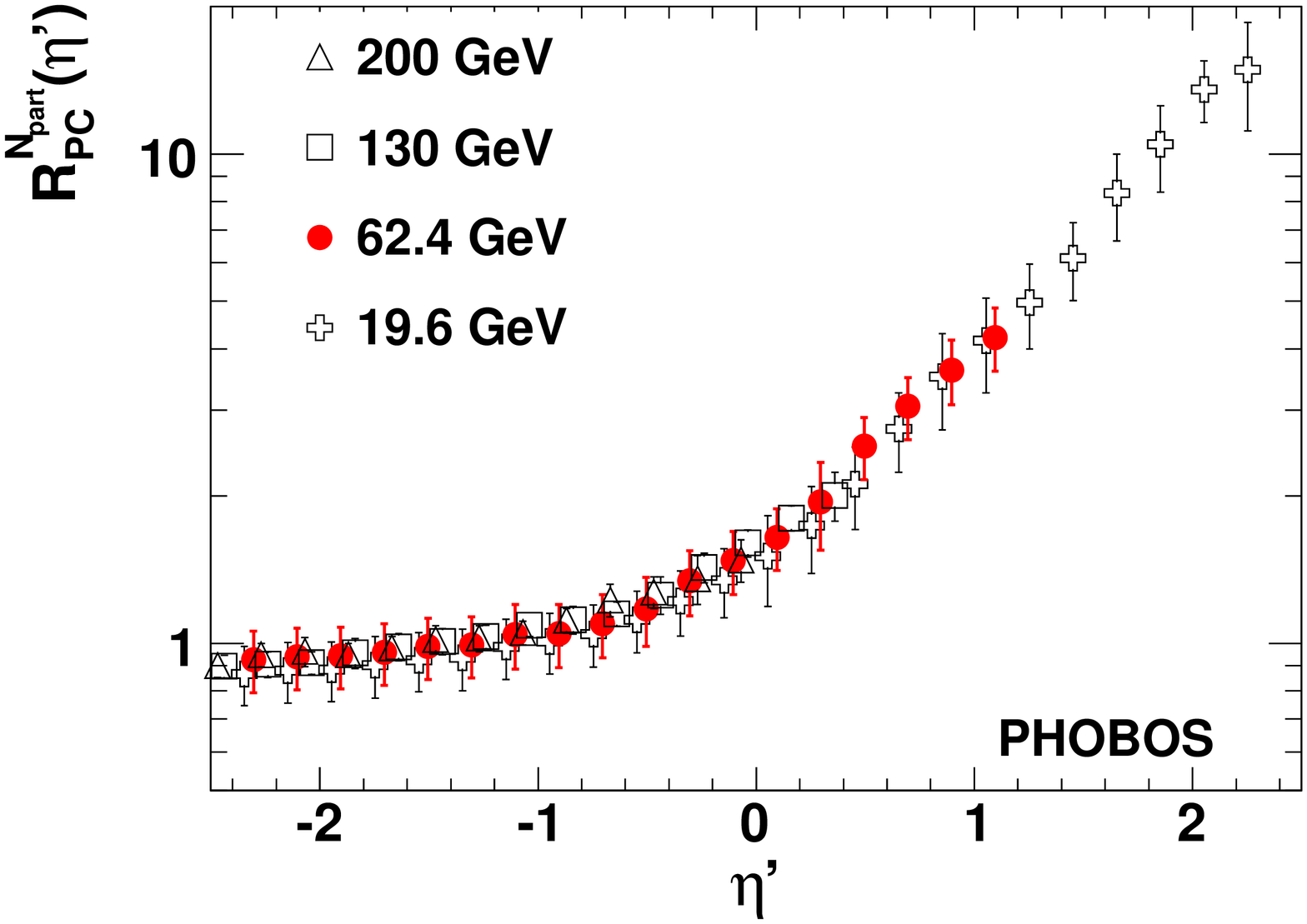}}
\caption{\label{aalimfrag} Left panel: distributions of pseudorapidity density of charged 
particles emitted in a) central, b) peripheral Au+Au collisions at various   
energies as a function of $\eta'=\eta-y_{\rm beam}$, normalized by the average number of 
participant nucleon pairs. Right panel: ratio of $\eta'$ distributions in peripheral and 
central Au+Au collisions, normalized to the number of participant pairs \cite{gunthertalk}.}
\vspace*{-3mm}
\end{figure}

In heavy ion collisions at RHIC energies, the BRAHMS and PHOBOS experiments have collected 
a large and systematic dataset on $dN/d\eta$ distributions, at 
$\sqrt{s_{_{\rm NN}}}$= 19.6, 62.4, 130 and 200 GeV energies in Au+Au 
\cite{phoboswhite,brahmsdndeta}, and at 62.4 and 200 GeV energies in Cu+Cu collisions 
\cite{gunthertalk}. Instead of a sizeable plateau at midrapidity, 
similarly to p+p and p+A collisions,
an extended longitudinal scaling region 
around $\eta'=0$ is observed, as shown in the left panel of Fig. 
\ref{aalimfrag}. One can visualize this scaling by imagining that the energies of the two 
colliding beams are adjustable separately. If one holds the energy of the first beam
constant while increasing the energy of the second beam (in which the low-x gluons become
more and more saturated), the particle density measured at a fixed $\eta$ increases only 
up to a limit, given by the limiting curve in Fig. \ref{aalimfrag}, where it saturates.
This longitudinal scaling holds for Au+Au events with the same collision 
centrality (same number of participants), but the limiting curve is different for central 
and for peripheral events. The ratio of $dN/d\eta'$ distributions in peripheral to central 
Au+Au collisions is found to be energy independent to a remarkable precision
over an order of magnitude in $\sqrt{s}$, as shown in the right panel of Fig. \ref{aalimfrag}.

\begin{figure}[t]
\centerline{
\vspace*{-8mm}
\includegraphics[width=70mm]{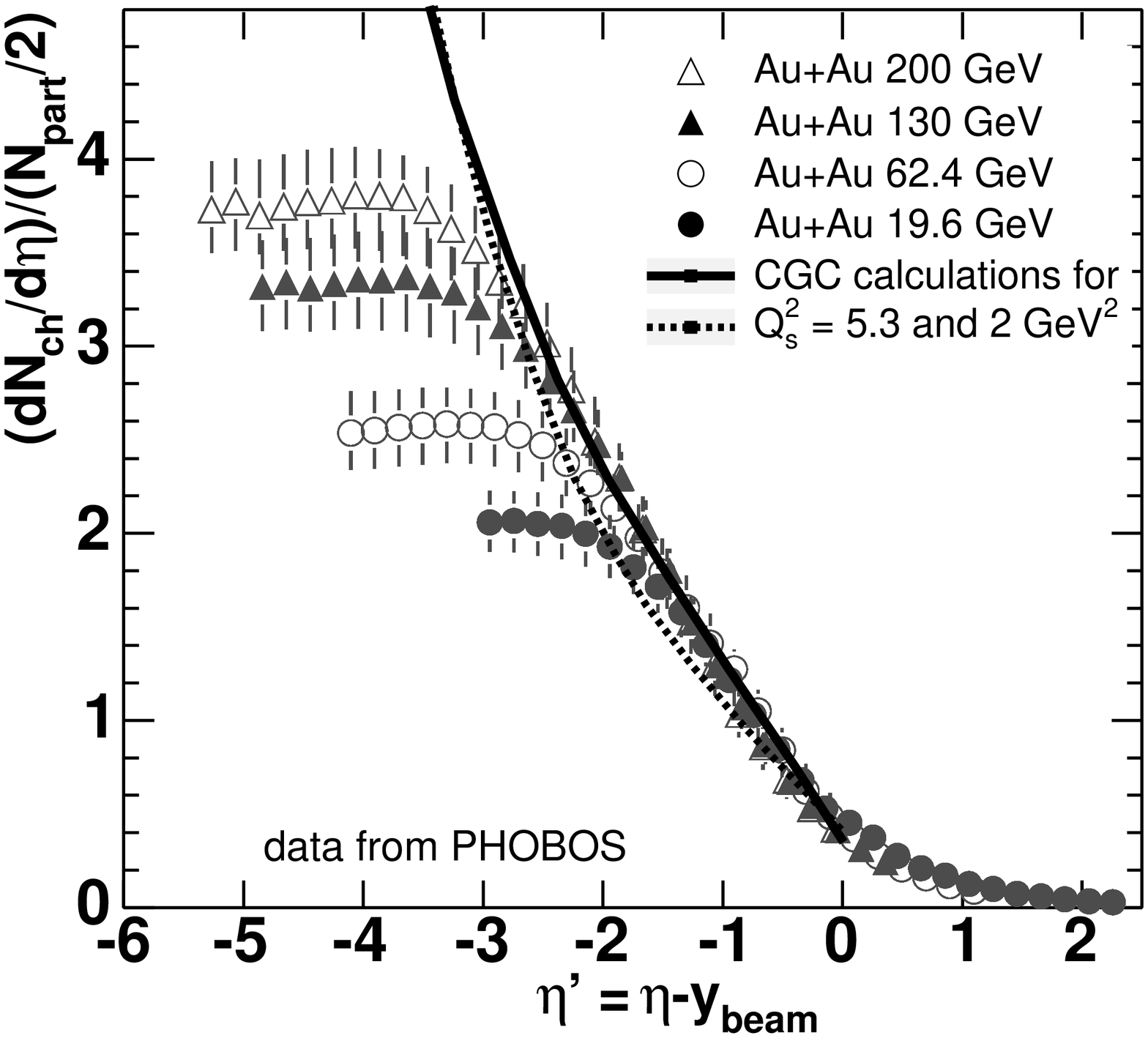}
\hspace{12mm}
\includegraphics[width=66mm]{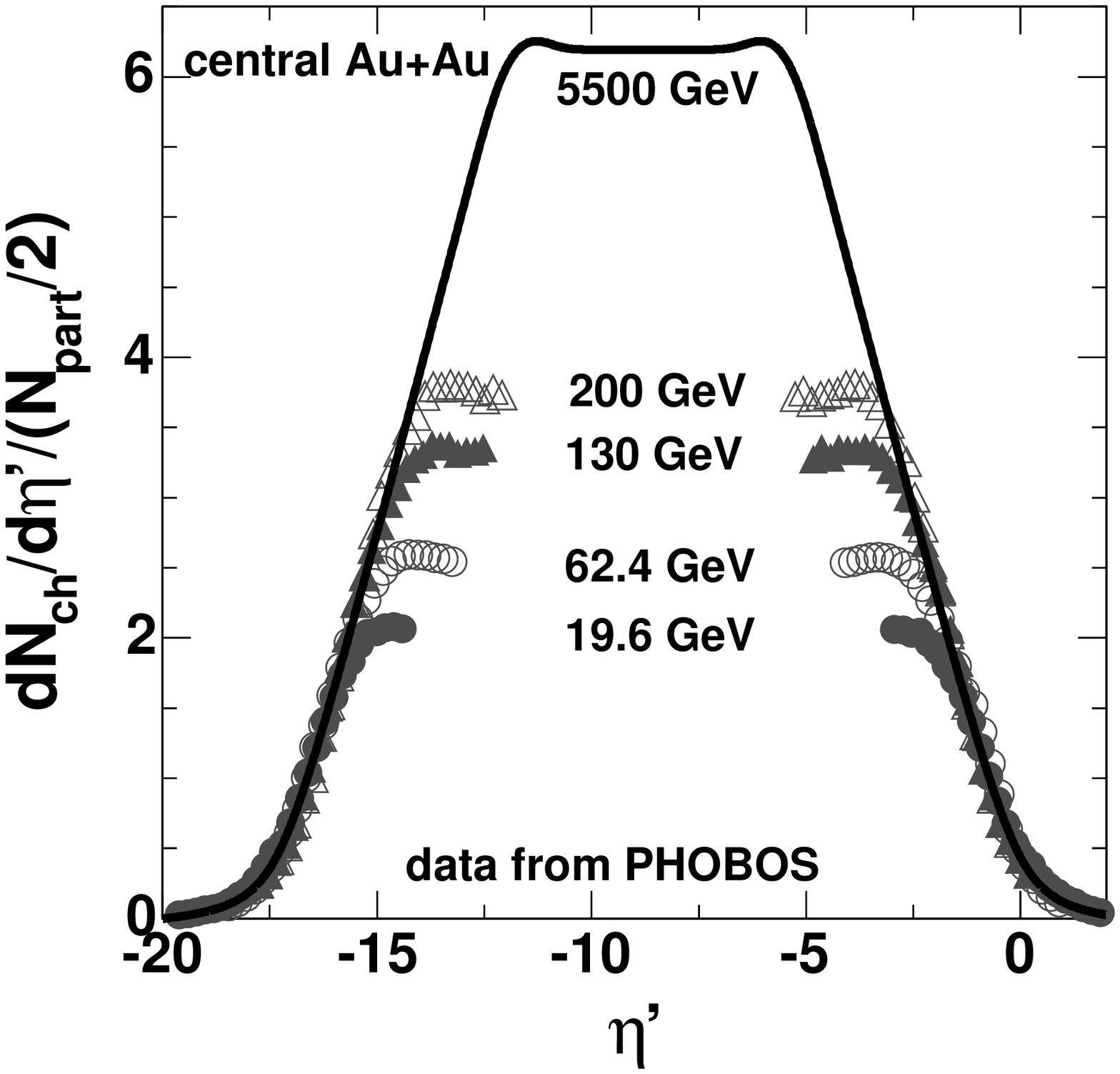}}
\caption{\label{lfpred} Left panel: $dN/d\eta'$ distributions 
of charged particles emitted in central Au+Au 
collisions at various energies as a function of $\eta'=\eta-y_{\rm beam}$, 
normalized by the average number of participant nucleon pairs, together with
CGC calculations \cite{jamalcgc}. Right 
panel: extrapolation to LHC energy based on {\it linear} limiting curve
and {\it logarithmic} rise of midrapidity density with $\sqrt{s}$ \cite{buszaschool}.}
\vspace*{-4mm}
\end{figure}       

Longitudinal scaling is compatible with the Color Glass Condensate picture.
The application of gluon saturation and standard parton fragmentation to
the high rapidity region leads to such a limiting curve \cite{jamalcgc}.  This
prediction is shown in the left panel of Fig. \ref{lfpred}, and is in good agreement
with the data.

Of course, the theoretical description breaks down at a certain $\eta'$ value,
as the probed $x_{\rm Bj}$ in the projectile increases. 
The centrality dependence of the $dN/d\eta$ distributions is also correctly described 
by the Color Glass Condensate model \cite{jamalcgc2}. 

The predictions of various model calculations for
$dN/d\eta$ at $\eta=0$ in central Pb+Pb collisions at the LHC ($\sqrt{s_{_{\rm 
NN}}}$=5500 GeV) range between 1000 and 7000.
Based on the simple assumptions that the limiting curve of the 
$dN/d\eta'$ distributions is linear (as seems to be the case at RHIC energies)
and that the observed logarithmic increase of $dN/d\eta|_{\eta=0}$ with $\sqrt{s_{_{\rm NN}}}$
continues to be valid \cite{phobosprl}, one can estimate the central 
$dN/d\eta$ to be about 1100 and the total number of charged particles created to 
be about 14000 at the LHC \cite{buszaschool}, see the right panel of Fig. \ref{lfpred}. 

\subsection{Identified particles}

\begin{figure}[t]
\centerline{
\vspace*{-8mm}
\includegraphics[width=48mm]{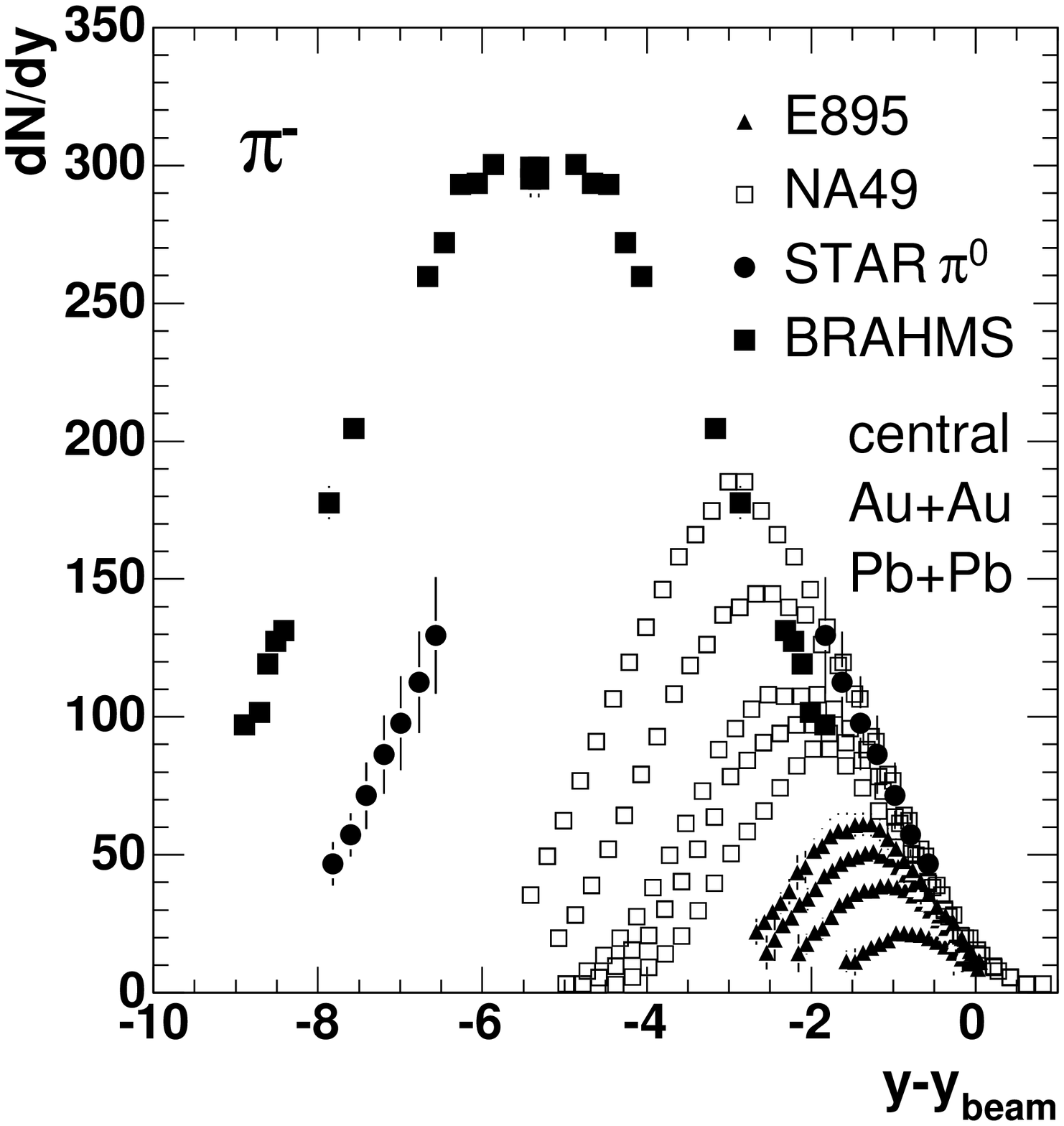}
\hspace{2.8mm}
\includegraphics[width=53mm]{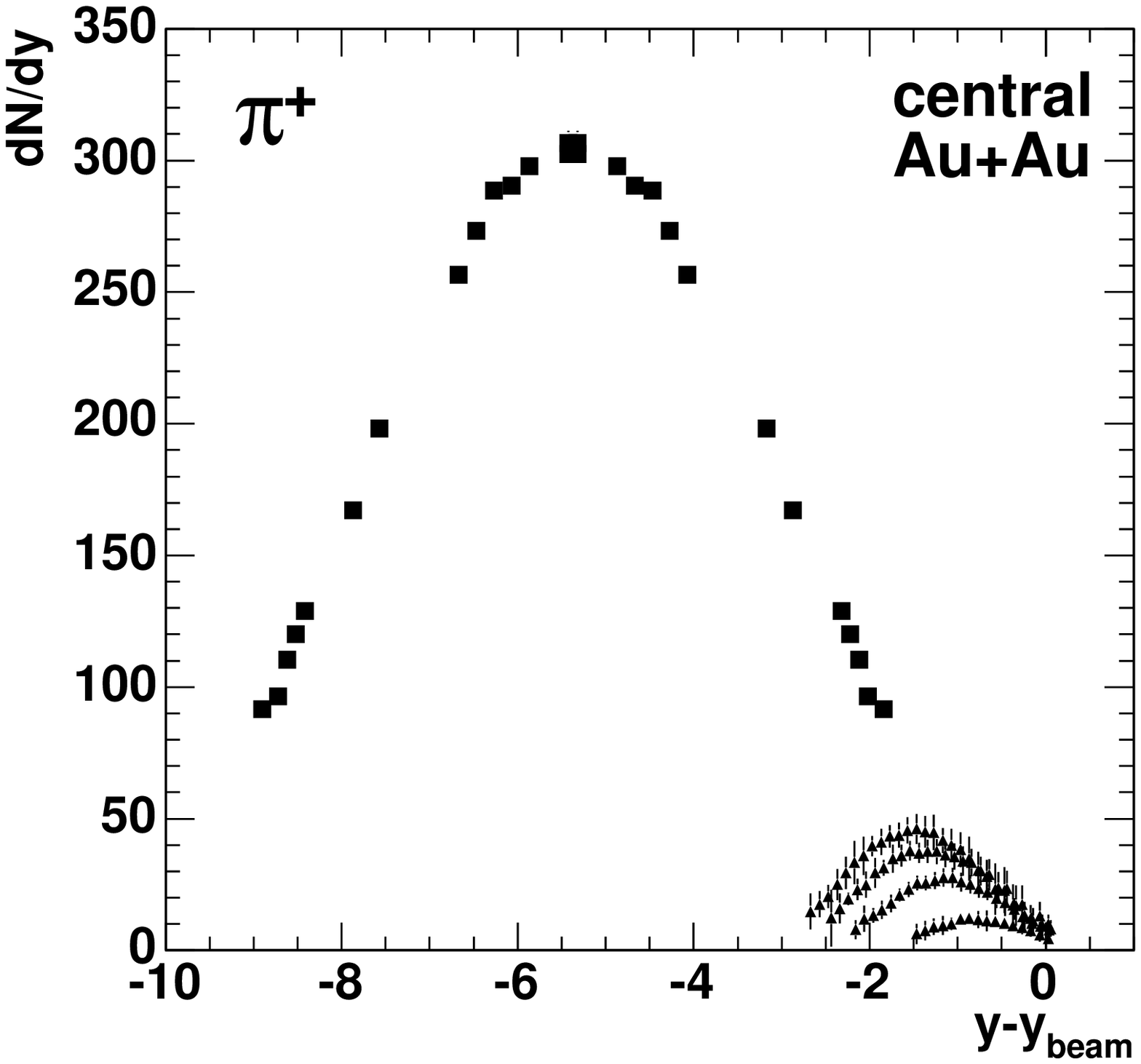}
\hspace{4.3mm}
\includegraphics[width=52mm]{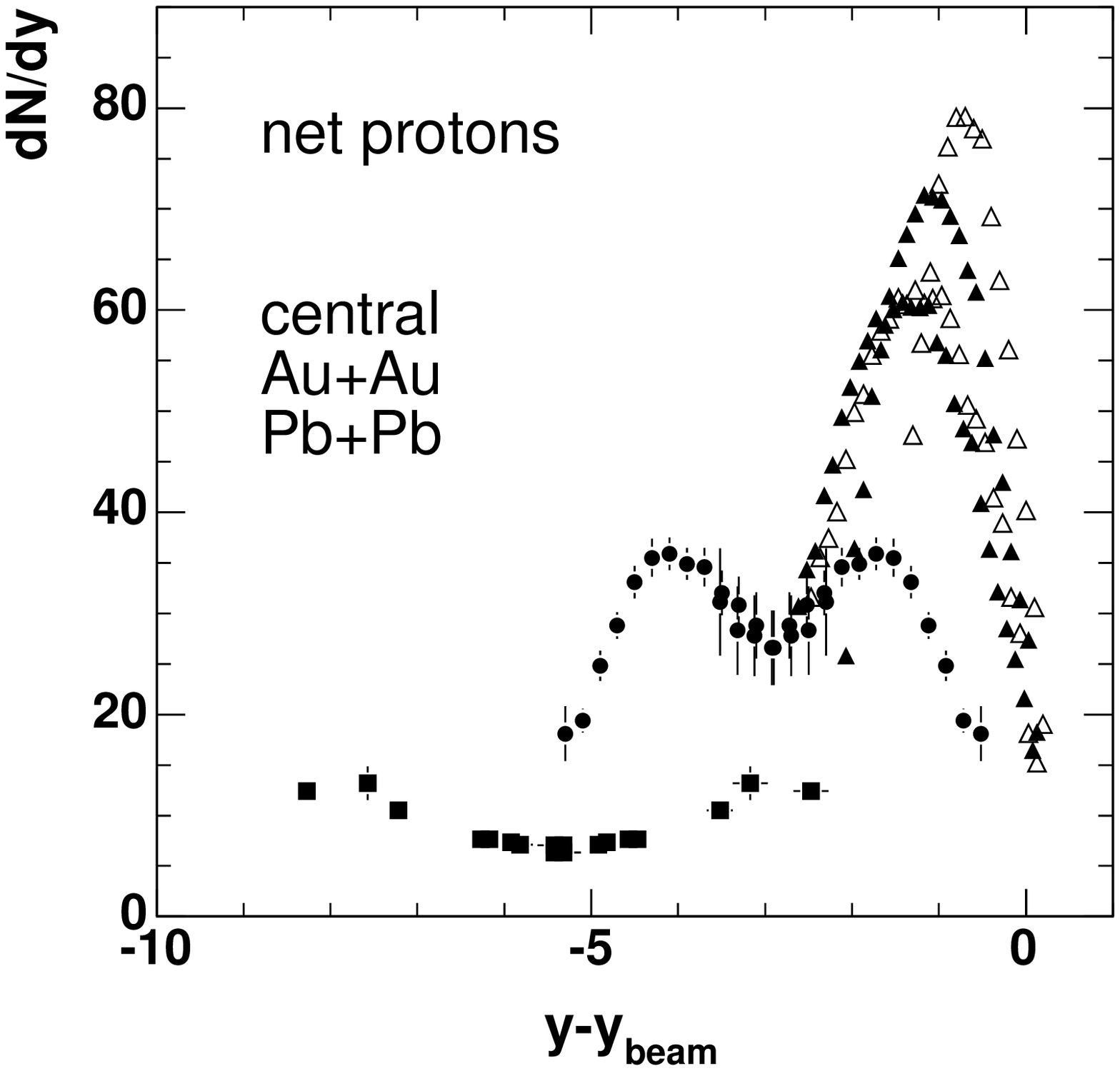}}
\caption{\label{identified} Rapidity ($y'$=$y$-$y_{\rm beam}$) distributions of  
$\pi^-$ and $\pi^+$ particles and net protons 
($p-\overline{p}$) produced in central Au+Au (Pb+Pb) collisions
at $\sqrt{s_{_{\rm NN}}}=$ 2.63, 3.28, 3.84, 4.29
\cite{agspionsprotons}, 6.27, 7.62, 8.76, 12.32, 17.27
\cite{na49piminus,na49netprotons}
and 200 GeV \cite{brahmspions,brahmsnetprotons}, and $\pi^0$-s
at 62.4 GeV \cite{starphotonpizero}.}
\vspace*{-4mm}
\end{figure}

The scaling properties discussed so far are more surprising if one 
considers the different production mechanism of various particle 
species. The abundance of the lightest mesons indicate the 
amount of entropy created (in a thermodynamical picture).
The amount of net baryons ($B-\overline{B}$) however is a conserved 
quantity, and the amount of stopping of the initial baryons in the 
colliding nuclei gives rise to the energy necessary for meson production 
as well as baryon pair production. 

There are extensive measurements of the rapidity distributions
of charged pions 
\cite{agspionsprotons,na49piminus,brahmspions}, protons 
and antiprotons \cite{agspionsprotons,na49netprotons,brahmsnetprotons} in 
heavy ion collisions. As shown in Fig. \ref{identified} for
collision energies between 2.63 and 200 GeV per nucleon pair, 
the charged pion production, observing it in the rest frame of one of the 
projectiles, shows a similar scaling behaviour as the charged hadrons
as a whole.
The recent measurement of the neutral pion yield at high rapidities
at 62.4 GeV \cite{starphotonpizero} follows the scaling observed for 
the charged pions rather precisely.
A similar scaling of net proton distributions is not expected because of 
baryon conservation and a widening of the rapidity gap with increasing 
energy. The right panel of Fig. \ref{identified} is a compilation of 
$dN/dy$ distributions of protons (below $\sqrt{s_{_{\rm NN}}}$=5 GeV)
and net protons (at $\sqrt{s_{_{\rm NN}}}$=17.27 and 200 GeV).

The natural question arising is whether this longitudinal scaling
at high rapidity is a fundamental feature of particle production,
or a remarkable coincidence, considering that many effects (baryon 
stopping, gluon saturation, collective effects, parton fragmentation etc.) 
play a role at the same time. Naturally, the scaling behaviour cannot hold 
equally precisely for all charged particles and for only mesons,
since the dynamics of baryon stopping and production is fundamentally 
different. This could perhaps explain why the limiting curve for charged 
particles appears to be centrality dependent, as opposed to photons - 
originating mostly from $\pi^0$ decays \cite{starphotonpizero}).

\subsection{Feynman-x scaling of various final states}

\begin{figure}[t]
\centerline{
\vspace*{-8mm}
\includegraphics[width=67mm]{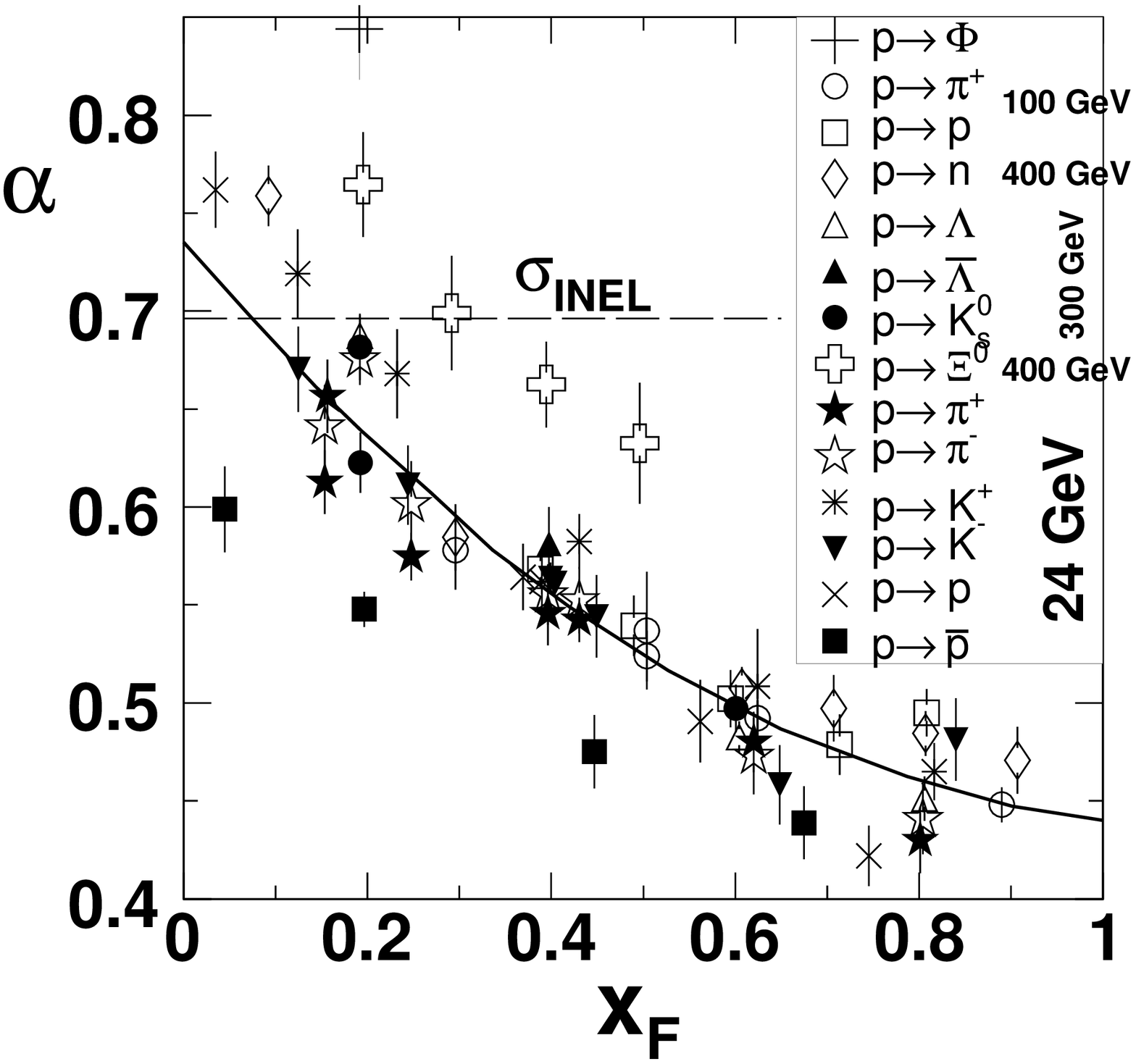}
\includegraphics[width=91mm]{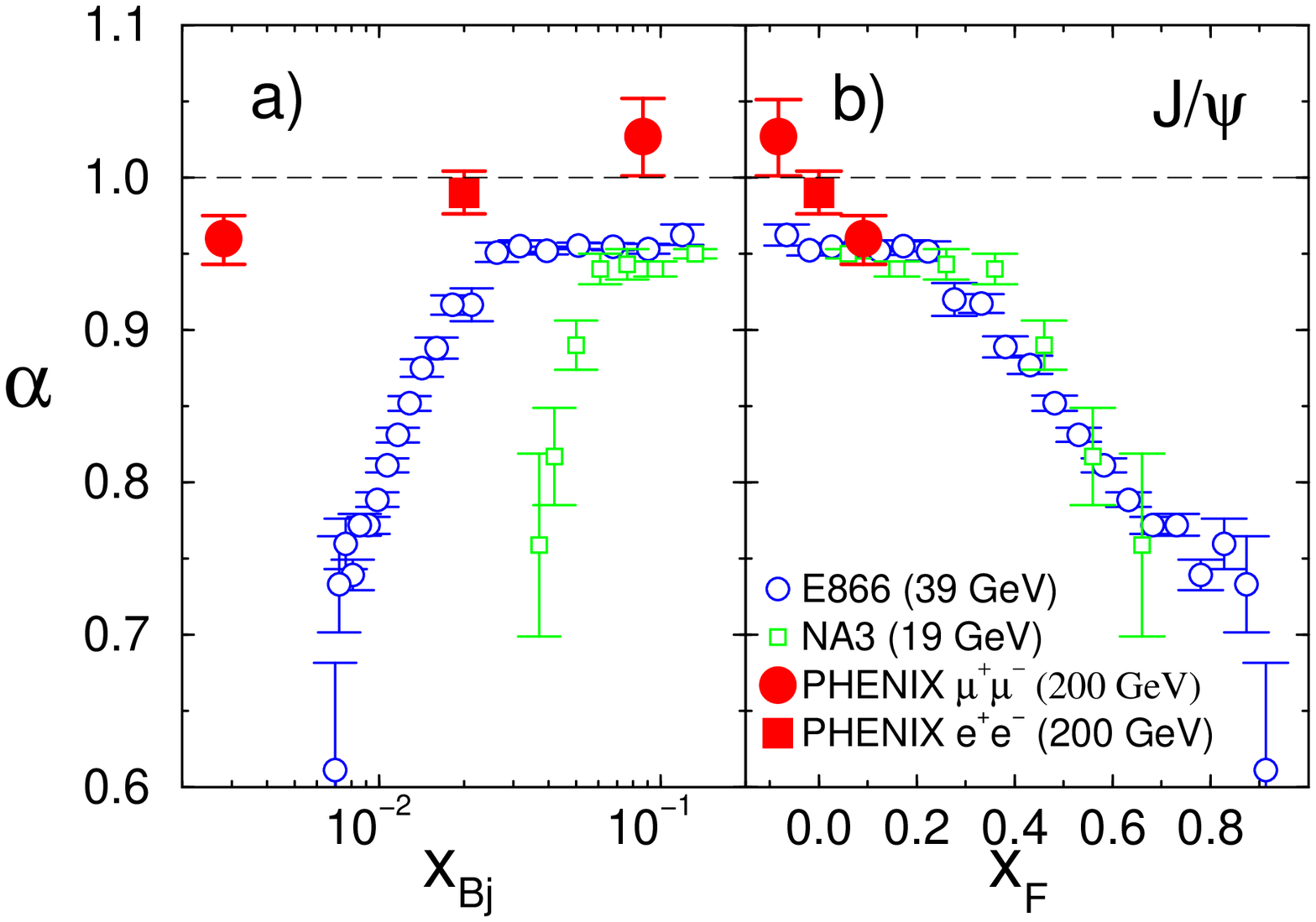}}
\caption{\label{baryon2} Scaling of the production of
$\Phi,\pi^+,\pi^-,p,\overline{p},n,\Lambda,\overline{\Lambda},K^0_s,\Xi^0,K^+,K^-$
and $J/\Psi$ particles as a function of $x_F$ (left and right panels) and $x_{\rm Bj}$ in the target
nucleus (middle panel) in p+A \cite{buszaxf} and d+Au \cite{phenixjpsi}
collisions at $\sqrt{s_{_{\rm NN}}}$=19, 39 and 200 GeV.}
\vspace*{-4mm}
\end{figure}

A frequently used way of quantifying longitudinal scaling and target size 
dependence of particle production in p+A collisions is to study the 
$\alpha$ exponent in the relation $\sigma_{\rm pA}=\sigma_0A^\alpha$.
This is done by measuring the $\sigma_{\rm pA}$ production cross sections as a function 
of the target nuclear mass, $A$, and as a function of the Feynman-x 
variable. The left panel of Fig. \ref{baryon2}
is a compilation of data showing various final states and collision energy in p+A collisions
\cite{buszaxf}. This shows that the attenuation as a 
function of $x_F$ is remarkably uniform within the experimental 
uncertainties. A more recent measurement of $J/\Psi$ mesons in d+Au 
collisions at $\sqrt{s_{_{\rm NN}}}$=200 GeV \cite{phenixjpsi} is
compared to lower energy data (right panel of Fig. \ref{baryon2}) and 
also exhibits the $x_F$ scaling with energy. However, the same scaling does 
not hold if the same data is plotted as a function of the $x_{\rm Bj}$ 
variable (referring to the Au nucleus. 
This indicates that the way in which gluon 
saturation plays a role in the $J/\Psi$ production is strongly energy 
dependent (middle panel of Fig. \ref{baryon2}). 

In the following section more details about the 
centrality dependence of hadron production (and baryon stopping) at high 
rapidities are discussed.

\section{BARYON STOPPING AND NUCLEAR TRANSPARENCY}

It is apparent from the previous sections, that the dynamics of the 
energy loss by the ini\-ti\-al baryons in a heavy ion collision strongly 
influences particle production (at high ra\-pi\-di\-ties). Baryon energy loss
in the hot medium created in heavy ion collisions (A+A) and in cold 
nuclear matter (p+A, d+A) has been studied and compared 
extensively (e.g. \cite{buszaxf,gaborposterqm01,thesisg,thesisr}). In addition, 
baryons are observed to behave differently than mesons when studying the 
`jet quenching' of different species, leading to 
the baryon `anomaly'  at RHIC \cite{banomaly}. 

\begin{figure}[t]
\centerline{
\vspace*{-8mm}
\includegraphics[width=40mm]{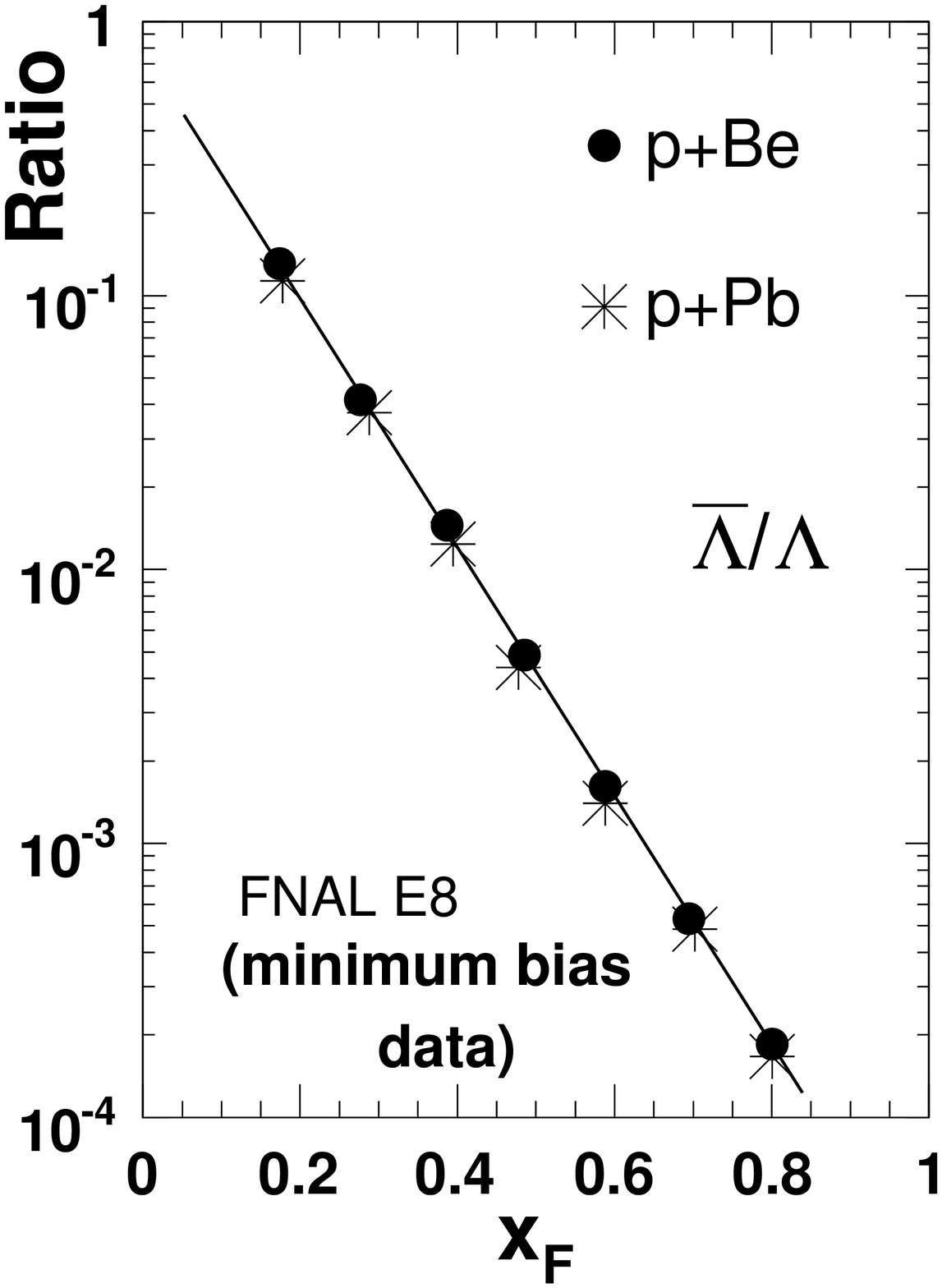}
\includegraphics[width=40mm]{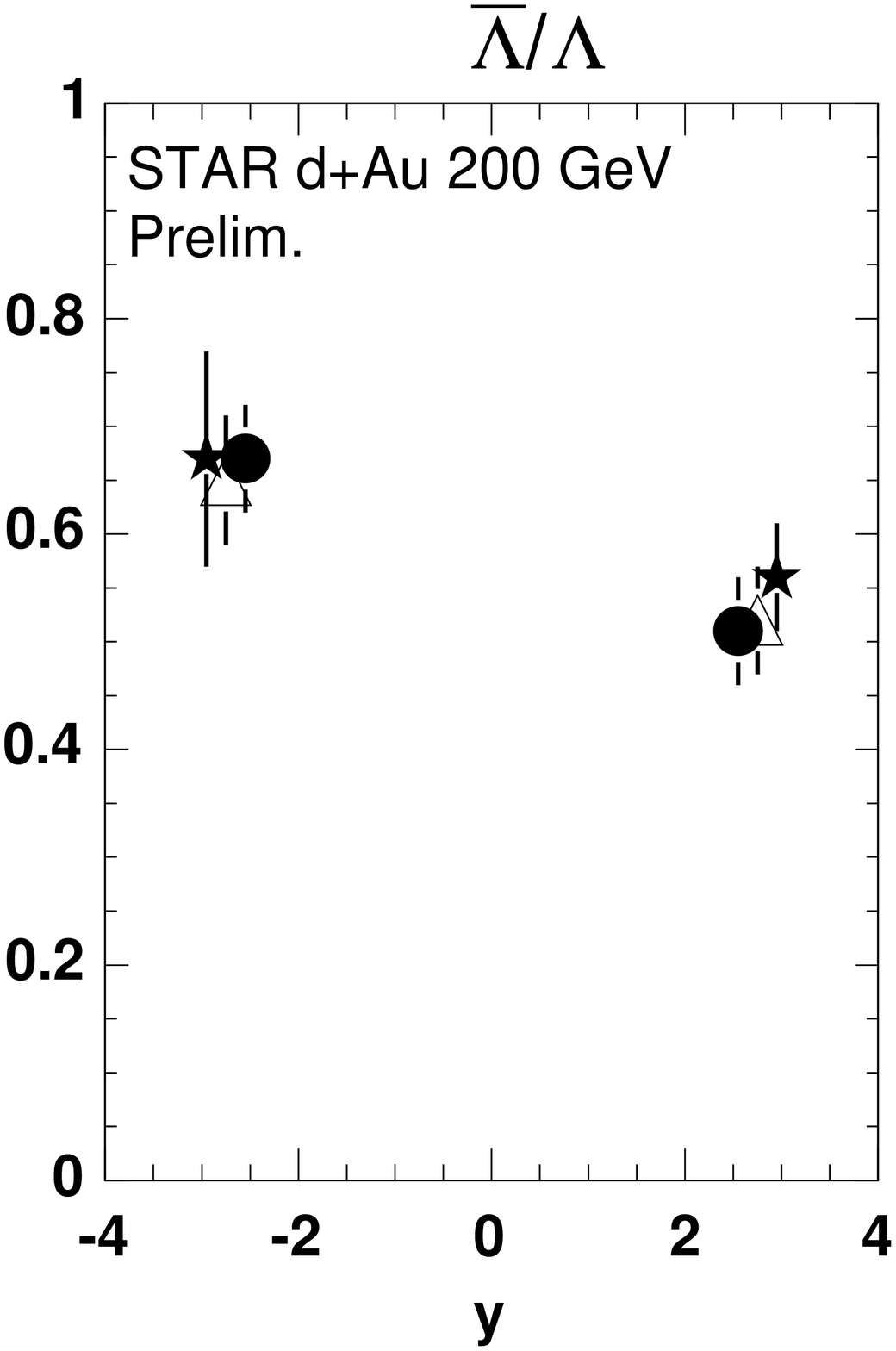}
\includegraphics[width=40mm]{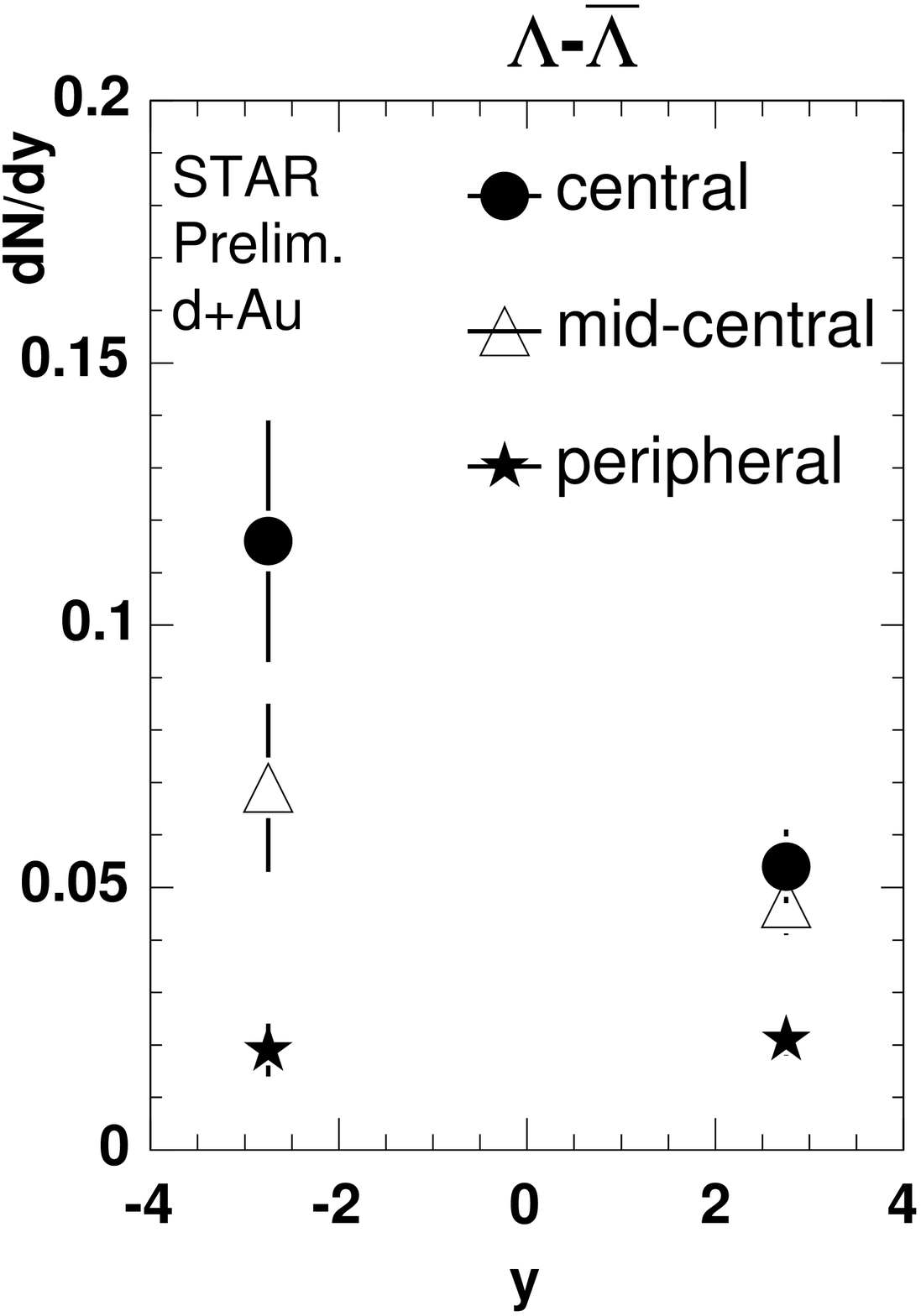}
\includegraphics[width=39mm]{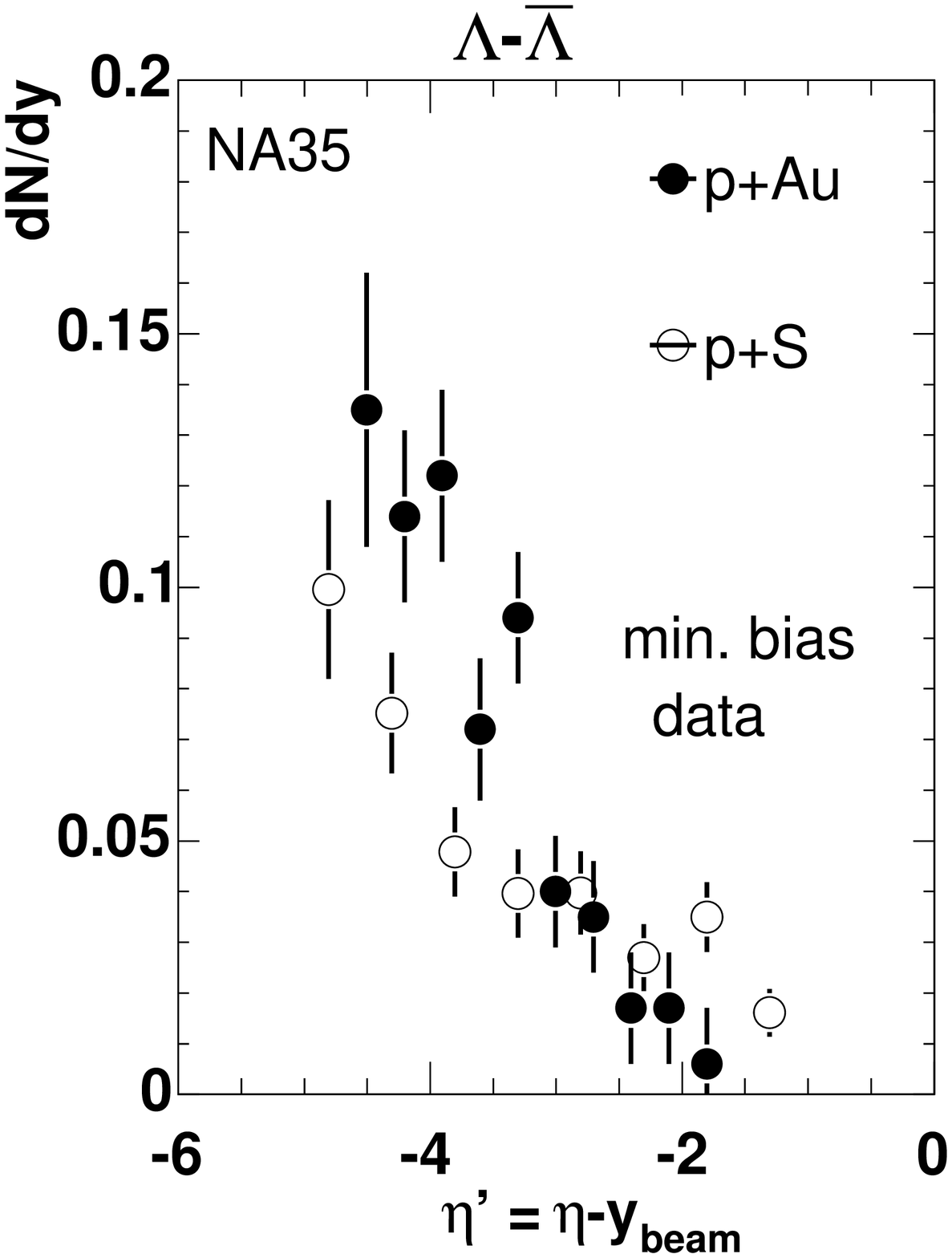}}
\caption{\label{baryon} $\Lambda/\overline{\Lambda}$ ratio vs. $x_F$
from \cite{llbar} in p+A collisions (300 GeV protons on fixed target); 
the same ratio in d+Au collisions at 
$\sqrt{s_{_{\rm NN}}}$=200 GeV from \cite{starlambda} (preliminary data); 
the net $\Lambda$ rapidity distribution from \cite{starlambda}; and in 
min. bias p+A collisions at the SPS \cite{na35lambdas}.}
\vspace*{-6.5mm}   
\end{figure}

Among the earlier puzzles on nuclear transparency is the 
independence of the $\Lambda/\overline{\Lambda}$ ratio on the size of 
the target nucleus (left panel of Fig. \ref{baryon}) seen in minimum bias p+A 
collisions at 300 GeV beam energy on fixed target \cite{llbar}. 
A complete attenuation in the cold nucleus, which would 
restrict particle production to the nuclear surface, may seem to be a 
plausible explanation. Preliminary data on $\Lambda$ hyperons at RHIC 
\cite{starlambda} show that this ratio is independent of centrality in 
d+Au collisions (second panel in Fig. \ref{baryon}). However, the 
$dN/dy$ yield of net lambdas ($\Lambda-\overline{\Lambda}$) strongly 
depends on target size and centrality, showing clear evidence of more baryon 
stopping for larger targets and central collisions, shown
in the two rightmost panels of Fig. \ref{baryon}
(200 GeV proton beam on fixed Au and S targets \cite{na35lambdas} 
and d+Au collisions at $\sqrt{s_{_{\rm NN}}}$=200 GeV \cite{starlambda}).
Similar observations have been made 
for the $p/\overline{p}$ ratio and for the net protons ($p-\overline{p}$), see
for example \cite{na35lambdas}. Interestingly, in d+Au and Au+Au collisions at 
$\sqrt{s_{_{\rm NN}}}$=200 GeV energy, the $p/\overline{p}$ ratio does not, or 
only weakly depends on centrality at midrapidity \cite{ppbarratio}, while
the amount of net protons is a strong (in fact {\it linear}) function of the 
number of participants in Au+Au collisions at $\sqrt{s_{_{\rm NN}}}$=62.4 and 
200 GeV \cite{conorsthesis}. Other studies of p+Pb and Pb+Pb collisions 
at SPS have also shown that while there is no complete attenuation for 
baryons, the net proton distributions have a very strong centrality 
dependence. Remarkably, the stopping is 
stronger in central p+Pb than central Pb+Pb collisions, 
even if samples with the same average
number of binary nucleon-nucleon collisions per participant are compared
\cite{gaborposterqm01,thesisg,thesisr}.

In {\it central} Au+Au collisions at $\sqrt{s_{_{\rm NN}}}$=200 GeV, the average rapidity 
loss of the incoming baryons was estimated to be about two units 
\cite{brahmsnetprotons}, and a similar value was obtained earlier for 
{\it minimum bias} p+A collisions \cite{buszaledoux}. Furthermore, baryon stopping (the rapidity 
distribution of 
$p-\overline{p}$) was found not to depend on the size of the projectile 
nucleus as long as it was smaller than the target nucleus and the 
collision was central \cite{na35lambdas}. 

The examples above show that it is not only necessary to understand
baryon stopping to separate it from other physically important phenomena.
To reach the correct conclusions, one must consider the relation of high-
and mid-rapidity physics, small and large colliding systems
and the centrality dependence of
the various observables, all at the same time. Studies of baryon 
stopping may also be able to address profound physical questions 
about the nature and carriers of baryon number. As an example, the HIJING 
B$\overline{\rm B}$ 2.0 model including {\it baryon junctions} can 
describe the rapidity dependence of {\it both} the $p/\overline{p}$ ratio 
{\it and} the net baryon density in 200 GeV Au+Au collisions \cite{vasile}.

\section{AZIMUTHAL ASYMMETRY OF HIGH-$\eta$ PARTICLE PRODUCTION}

\begin{figure}[t]
\centerline{
\vspace*{-8mm}
\includegraphics[width=85mm]{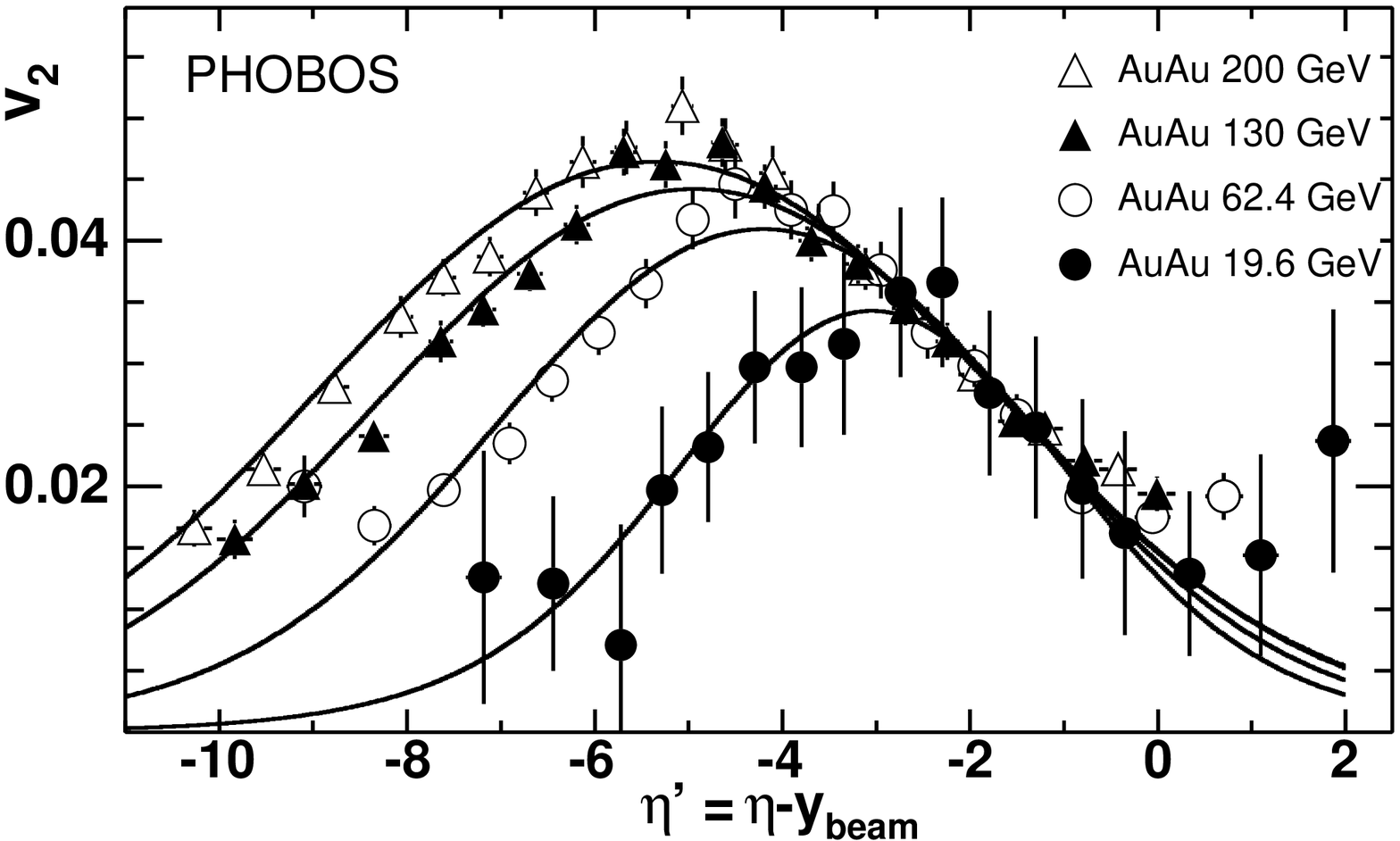}
\includegraphics[width=73.5mm]{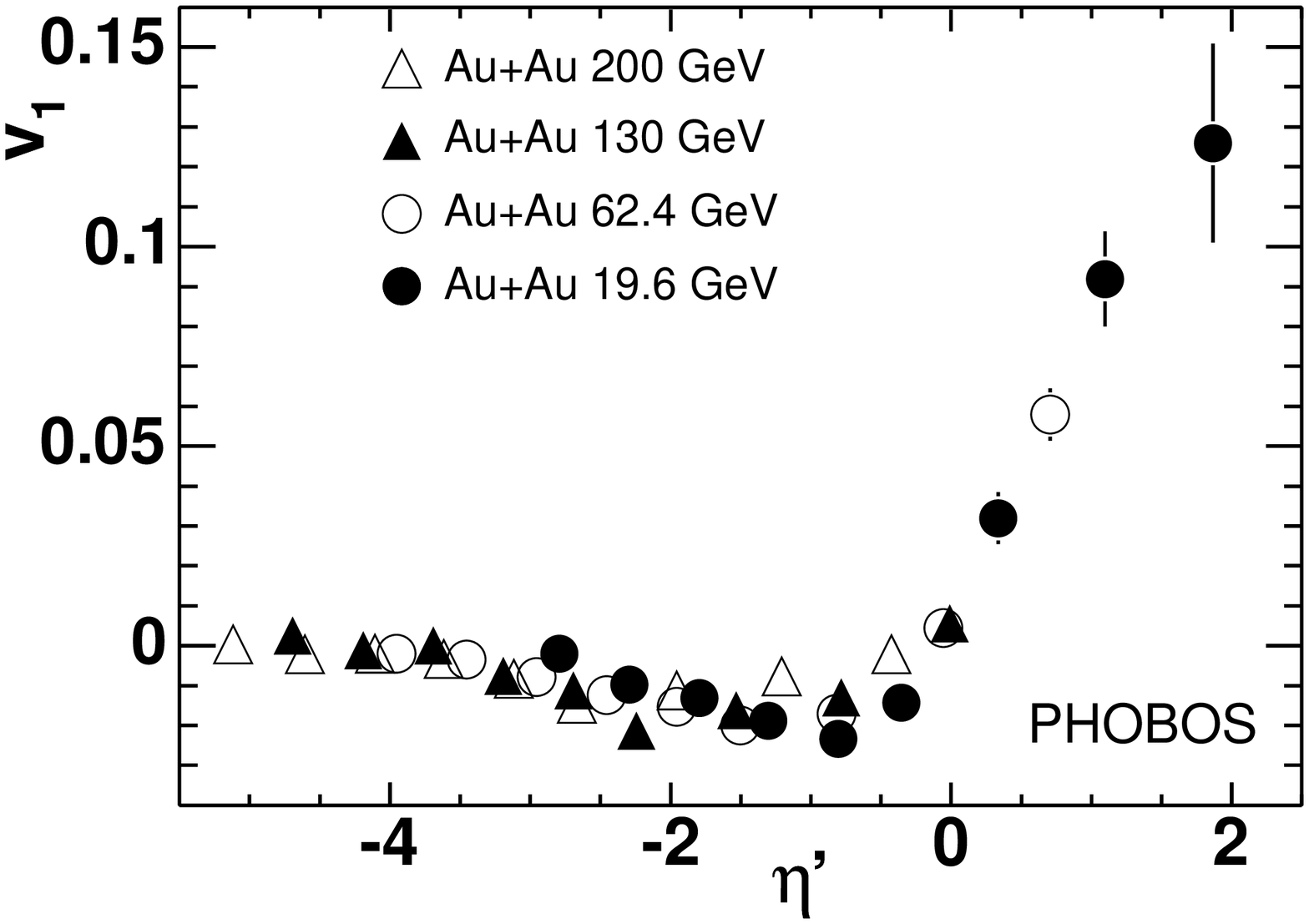}}
\caption{\label{flow} The $v_2$ measure of elliptic 
collective flow (left panel) and the $v_1$ measure of directed flow 
(right panel) in Au+Au collisions at $\sqrt{s}=$ 19.6, 62, 130 and 200 
GeV energy, as a function of $\eta'=\eta-y_{\rm beam}$ 
\cite{phobosv2,gunthertalk}. (The centrality is 0-40\% of the total inelastic cross section.)
Theoretical curves were made using the Buda-Lund model \cite{budalund}.} 
\vspace*{-4mm}
\end{figure}

One of the important `bulk features' of hadron production at high 
rapidities is the asymmetry of the azimuthal distribution of 
particles around the beam direction, relative to the reaction plane
(`flow'). This asymmetry can be elliptical,
characterized by the $v_2$ coefficient, as well as off-center or 
`directed', characterized by the $v_1$ coefficient of its 
Fourier-expansion.
This flow is a signature of collective, final state effects, as opposed to the 
gluon saturation at low $x_{\rm Bj}$, which is truly a feature of the 
{\it initial} 
state in the collision. Thus, the saturation picture alone cannot account 
for the large $v_2$ values observed in A+A collisions at RHIC. 
Furthermore, there is no reason to expect that similar longitudinal 
scaling would exist for the flow observables, as it did for the 
$dN/d\eta'$ distributions in that picture. 
In fact scaling is found to hold  \cite{gunthertalk,phobosv2},
as can be seen in Fig. \ref{flow}, both for the $v_2$ and the 
$v_1$ variables, in Au+Au collisions from 19.6 to 200 
GeV\footnote{Note that the strong $\eta$ dependence of $v_2$
was not observed by the BRAHMS experiment \cite{brahmsv2}.}. The 
elliptic flow data are well described by an analytical 
hydrodynamic model assuming ideal fluid dynamics \cite{budalund}.
This agreement illustrates how important {\it final} state 
interactions are for particle distributions at high rapidities.

\section{SUMMARY}

In summary, an extensive amount of data on hadron production at high
rapidities has been collected and is currently under analysis and
interpretation.  From these data only a few results could be selected for
discussion.
Longitudinal scaling of $dN/d\eta$ and $dN/dy$ distributions has been 
observed
as a function of energy for pions and charged hadrons but not for
protons.
Scaling features, the centrality dependence and $dN/d\eta$ 
distributions are successfully described by initial state gluon saturation 
models. However, baryon stopping (the influence of the valence 
structure of the projectile) also contributes to the final state hadron yields 
at high rapidity, following complicated dynamics having energy, centrality and 
target size dependence. Remarkably, another important `bulk feature' of 
particle production, the azimuthal asymmetry, 
which is not explainable by initial state effects alone, still shows 
the same kind of longitudinal scaling (energy independence) in heavy ion 
collisions. Theoretical progress in describing high $\eta$ (low-x) 
particle production in p+A and A+A collisions is rapidly progressing 
and generating a lot of interest. Nevertheless for a 
clean experimental isolation of initial state
modifications of the incoming projectile one might have to turn to more
differential measurements.  Ideas to separate the various physical
phenomena at high rapidities along with the new experiments at the LHC,
will guide our theoretical understanding.


\section{ACKNOWLEDGEMENTS}

I am indebted to W. Busza, D. R\"ohrich, M. 
Gyulassy, R. Venugopalan, K. Itakura, G. Roland, T. Cs\"org\H o, M. 
Csan\'ad, W. A. Zajc, B. Mohanty, C. H\"ohne, C. 
Blume, V. Topor-Pop, C. Loizides, R. Hollis, C. Henderson and F. Simon for the helpful 
discussions, and for providing me with 
their data. This work and participation at the Conference was possible due 
to the generous support from both of my Universities, from the Hungarian 
Scientific Research Fund under contracts T048898 and F049823, and from the 
J\'anos Bolyai Research Grant.

\begin{thebibliography}{9}
\bibitem{bjorken} J. D. Bjorken, Phys. Rev. D 27 (1983) 140.
\bibitem{phoboswhite} B. B. Back {\it et al.} [PHOBOS Collaboration], Nucl. Phys. A 757 (2005) 28.
\bibitem{brahmscgc} I. Arsene {\it et al.} [BRAHMS Collaboration], Phys. Rev. Lett. 93 (2004) 242303.
\bibitem{gyulassy} M. Gyulassy and L. McLerran, Nucl. Phys. A 750 (2005) 30.
\bibitem{jetq} K. Adcox {\it et al.} [PHENIX Collaboration], Phys. Rev. Lett. 88 (2002) 022301.
\bibitem{banomaly} K. Adcox {\it et al.} [PHENIX Collaboration], Phys. Rev. C 69 (2004) 024904.
\bibitem{vasile} V. Topor-Pop {\it et al.}, Phys. Rev. C 70 (2004) 064906.
\bibitem{starpmd} M. M. Aggarwal {\it et al.} [STAR Collaboration], Nucl. Instr. Meth. A499 (2003) 751.
\bibitem{starftpc} K. H. Ackermann {\it et al.} [STAR Collaboration], Nucl. Instr. Meth. A499 (2003) 713.
\bibitem{starfpd} J. Adams {\it et al.} [STAR Collaboration], Phys. Rev. Lett. 92 (2004) 171801.
\bibitem{phenixmuon} S. S. Adler {\it et al.} [PHENIX Collaboration], Phys. Rev. Lett. 94 (2005) 082302.
\bibitem{brahmsexp} M. Adamczyk {\it et al.} [BRAHMS Collaboration], Nucl. Instr. Meth. A 499 (2003) 437.
\bibitem{phobosnim} B. B. Back {\it et al.} [PHOBOS Collaboration], Nucl. Inst. Meth. A 499 (2003) 603.
\bibitem{adamtalk} A. Trzupek (for the PHOBOS Collaboration), these Proceedings.
\bibitem{murraytalk} M. Murray (for the CMS Collaboration), these Proceedings.
\bibitem{benecke} J. Benecke {\it et al.}, Phys. Rev. 188 (1969) 2159.
\bibitem{pplimfrag} F. Abe {\it et al.} [CDF Collaboration], Phys. Rev. D 41 (1990) 2330.\\
                   R. E. Ansorge {\it et al.} [UA5 Collaboration], Z. Phys. C 43 (1989) 75.\\
                   W. Thome {\it et al.} [ISR], Nucl. Phys. B 129 (1977) 365.
\bibitem{oldpa}
J. E. Elias {\it et al.} [E178 Experiment], Phys. Rev. D 22 (1980) 13.\\
I. Otterlund {\it et al.}, Nucl. Phys. B 142 (1978) 445. and references therein\\
S. Fredriksson {\it et al.} Phys. Rept. 144 (1987) 187. and references therein
\bibitem{phobosdAumult} B. B. Back {\it et al.} [PHOBOS Collaboration], {\it arXiv:nucl-ex/0409021} (2004).
\bibitem{brahmsdau} I. Arsene {\it et al.} [BRAHMS Collaboration], Phys. Rev. Lett. 94 (2005) 032301.
\bibitem{gunthertalk} G. Roland (for the PHOBOS Collaboration), these Proceedings.
\bibitem{brahmsdndeta} I. G. Bearden {\it et al.} [BRAHMS Collaboration], Phys. Rev. Lett. 88 (2002) 202301.
\bibitem{jamalcgc} J. Jalilian-Marian, Phys. Rev. C 70 (2004) 027902.
\bibitem{jamalcgc2} J. Jalilian-Marian, J. Phys. G 30 (2004) S751.
\bibitem{phobosprl} B. B. Back {\it et al.} [PHOBOS Collaboration], Phys. Rev. Lett. 88 (2002) 22302.
\bibitem{buszaschool} W. Busza, Acta Phys. Polon. B 35 (2004) 2873.
\bibitem{agspionsprotons} J. L. Klay {\it et al.} [E895 Collaboration], Phys. Rev. C 68 (2003) 054905.
\bibitem{na49piminus} S. V. Afanasiev {\it et al.} [NA49 Collaboration], Phys. Rev. C 66 (2002) 054902.
\bibitem{brahmspions} I. G. Bearden {\it et al.} [BRAHMS Collaboration], Phys. Rev. Lett. 94 (2005) 162301.
\bibitem{na49netprotons} H. Appelshauser {\it et al.} [NA49 Collaboration], Phys. Rev. Lett. 82 (1999) 2471.
\bibitem{brahmsnetprotons} I. G. Bearden {\it et al.} [BRAHMS Collaboration], Phys. Rev. Lett. 93 (2004) 102301.
\bibitem{starphotonpizero} J. Adams {\it et al.} [STAR Collaboration], Phys. Rev. Lett. 95 (2005) 062301.
\bibitem{phenixjpsi} S. S. Adler {\it et al.} [PHENIX Collaboration], {\it arXiv:nucl-ex/0507032} (2005).
\bibitem{buszaxf} W. Busza, Nucl. Phys. A 544 (1992) 49c.
\bibitem{gaborposterqm01} G. I. Veres, poster at the Quark Matter 2001 Conference
\bibitem{thesisg} G\'abor I. Veres, {\it Ph.D. Thesis}, ELTE, Budapest (2001).
\bibitem{thesisr} Andrzej Rybicki, {\it Ph.D. Thesis}, IFJ, Krak\' ow (2002).
\bibitem{llbar} P. Skubic {\it et. al}, Phys. Rev. D 18 (1978) 3115.
\bibitem{starlambda} F. Simon [for the STAR Collaboration], J. Phys. G 31 (2005) S1065.
\bibitem{na35lambdas} T. Alber {\it et al.} [NA35 Collaboration], Eur. Phys. J. C 2 (1998) 643.
\bibitem{ppbarratio} B. B. Back {\it et al.} [PHOBOS Collaboration], Phys. Rev. C 70 (2004) 011901(R).
\bibitem{conorsthesis} C$\underline{\rm o}$nor Henderson, {\it Ph.D. thesis}, Mass. Inst. Tech., Cambridge (2005) Fig. 6-8.
\bibitem{buszaledoux} W. Busza and R. Ledoux, Ann. Rev. Nucl. Part. Sci. 38 (1988) 119.
\bibitem{phobosv2} B. B. Back {\it et al.} [PHOBOS Collaboration], Phys. Rev. Lett. 94 (2005) 122303.
\bibitem{brahmsv2} H. Ito (for the BRAHMS Collaboration), these Proceedings.
\bibitem{budalund} M. Csan\'ad, T. Cs\"org\H o, B. L\"orstad and A. Ster, {\it arXiv:nucl-th/0510027} and {\it 0509106}.
\end{thebibliography}
\end{document}